\pdfoutput=1 
\documentclass{JINST}

\usepackage{multirow}

\title{Development of a 30~cm-cube Electron-Tracking
Compton Camera for the SMILE-II Experiment}

\author{Y.~Mizumura$^{a,b}$\thanks{Corresponding author.},
T.~Tanimori$^{b}$,
H.~Kubo$^{b}$,
A.~Takada$^{b}$,
J.~D.~Parker$^{b}$,
T.~Mizumoto$^{c}$,
S.~Sonoda$^{d}$,
D.~Tomono$^{b}$,
T.~Sawano$^{b}$,
K.~Nakamura$^{b}$,
Y.~Matsuoka$^{b}$,
S.~Komura$^{b}$,
S.~Nakamura$^{b}$,
M.~Oda$^{b}$,
K.~Miuchi$^{e}$,
S.~Kabuki$^{f}$,
Y.~Kishimoto$^{g}$,
S.~Kurosawa$^{h}$,
and~S.~Iwaki$^{b}$\\
\llap{$^a$}Unit of Synergetic Studies for Space, Kyoto University, Kyoto, Kyoto, 606-8502, Japan\\
\llap{$^b$}Department of Physics, Kyoto University, Kyoto, Kyoto, 606-8502, Japan\\
\llap{$^c$}Research Institute for Sustainable Humanosphere, Kyoto University, Uji, Kyoto, 611-0011, Japan\\
\llap{$^d$}Advanced Biomedical Engineering Research Unit, Kyoto University, Kyoto, Kyoto, 606-8501, Japan\\
\llap{$^e$}Department of Physics, Kobe University, Kobe, Hyogo, 657-8501, Japan\\
\llap{$^f$}Department of Radiation Oncology, Tokai University, Isehara, Kanagawa, 259-1193, Japan\\
\llap{$^g$}Radiation Science Center, KEK, Tsukuba, Ibaraki, 305-8501, Japan\\
\llap{$^h$}Institute for Materials Research, Tohoku University, Sendai, Miyagi, 980-8577, Japan\\
E-mail: \email{mizumura@cr.scphys.kyoto-u.ac.jp}}

\abstract{
To explore the sub-MeV/MeV gamma-ray window for astronomy,
we have developed the Electron-Tracking Compton Camera (ETCC),
and carried out the first performance test
in laboratory conditions using several gamma-ray sources
in the sub-MeV energy band.
Using a simple track analysis
for a quick first test of the performance,
the gamma-ray imaging capability was demonstrated with clear images and
5.3 degrees of angular resolution measure (ARM) measured at 662~keV.
As the greatest impact of this work,
a gamma-ray detection efficiency on the order of $10^{-4}$
was achieved at the sub-MeV gamma-ray band,
which is one order of magnitude higher than our previous experiment.
This angular resolution and detection efficiency enables us
to detect the Crab Nebula at the $5\sigma$ level
with several hours observation at balloon altitude in middle latitude.
Furthermore,
good consistency of efficiencies between this performance test
and simulation including only physical processes is very important;
it means we achieve nearly $100\%$ detection of
Compton recoil electrons and means that our predictions of performance
enhancement resulting from future upgrades are more realistic.
We are planning to confirm the imaging capability of the ETCC
by observation of celestial objects in the SMILE-II
(Sub-MeV gamma ray Imaging Loaded-on-balloon Experiment II).
The SMILE-II and following SMILE-III project
will be an important key of sub-MeV/MeV gamma-ray astronomy.}


\begin{document}
\section{Introduction}
The Sub-MeV/MeV gamma-ray window is one of the most interesting frontiers
for both of multi-wavelength and nuclear gamma-ray astrophysics.
There still remain many unobserved celestial objects
such as black hole candidates, gamma-ray bursts (GRBs),
supernove remnants (SNRs), gamma-ray pulsars,
active galactic nuclei (AGNs), and so on.
However, sufficient observation has not yet been achieved due to
difficulties of gamma-ray imaging and rejection of huge backgrounds.
Until now, only several tens of celestial objects emitting MeV gamma rays
have been reported by COMPTEL~\cite{comptel-catalog},
whereas {\it Fermi}-LAT found 1873 sources emitting sub-GeV/GeV gamma rays
during the first 24 months of the all-sky survey~\cite{fermi-catalog}.

In order to explore sub-MeV gamma-ray astronomy,
we have developed an Electron-Tracking Compton Camera (ETCC)
consisting of a gaseous Time Projection Chamber (TPC)
and pixel scintillator arrays (PSAs).
In comparison with a COMPTEL-type legacy Compton camera,
the ETCC measures three additional physical parameters
by obtaining the three dimensional track of the Compton recoil electron.
Knowing the direction of the recoil electron makes it possible to reconstruct
the direction of the incident gamma-ray event by event,
and the energy loss rate (dE/dx) of the charged particle
provides efficient background rejection.
%

Previously, we carried out the 
first balloon-borne experiment (SMILE-I) in 2006 using
a small size ETCC with a $10\times10\times15$ cm$^3$ TPC,
and we successfully observed diffuse cosmic and atmospheric gamma-ray spectra
with efficient background suppression~\cite{takada-11}.
To improve gamma-ray sensitivity,
we had developed and tested a prototype ETCC
with a (30~cm)$^3$ size TPC~\cite{ueno-12}.
More recently, we have started construction of
the next balloon flight ETCC system (SMILE-II) with a (30~cm)$^3$ TPC and 
completely upgraded the data acquisition system (DAQ).
The SMILE-II experiment is aimed to demonstrate
the gamma-ray imaging capability of the ETCC
by observing bright celestial gamma-ray point sources
such as the Crab Nebula or Cygnus X-1.
We are now planning one day balloon flights
at middle latitude in the northern hemisphere.
After the completion of the SMILE-II experiment,
our system will be upgraded for
balloon flights around the polar regions, SMILE-III.
These experiments are milestones to a satellite-based all-sky survey
for an unexplored frontier of astronomy.
%
%
%
Here, we report results of the first performance test
of the current SMILE-II ETCC with a simple analysis.

\section{Detectors: Electron-Tracking Compton Camera}

\begin{figure}[tbp]
  \centering
  \begin{tabular}{ccc}
    \includegraphics[width=.38\textwidth]{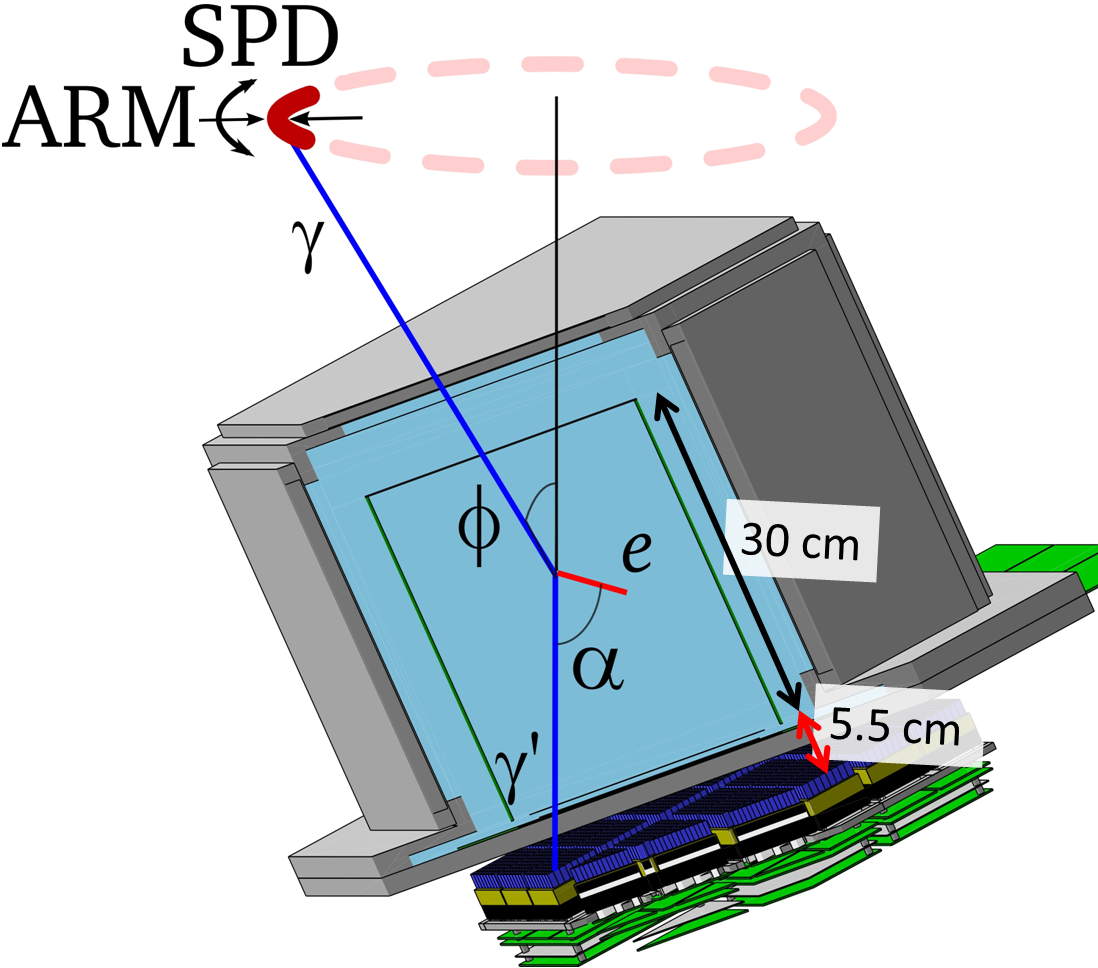} &
    \includegraphics[width=.17\textwidth]{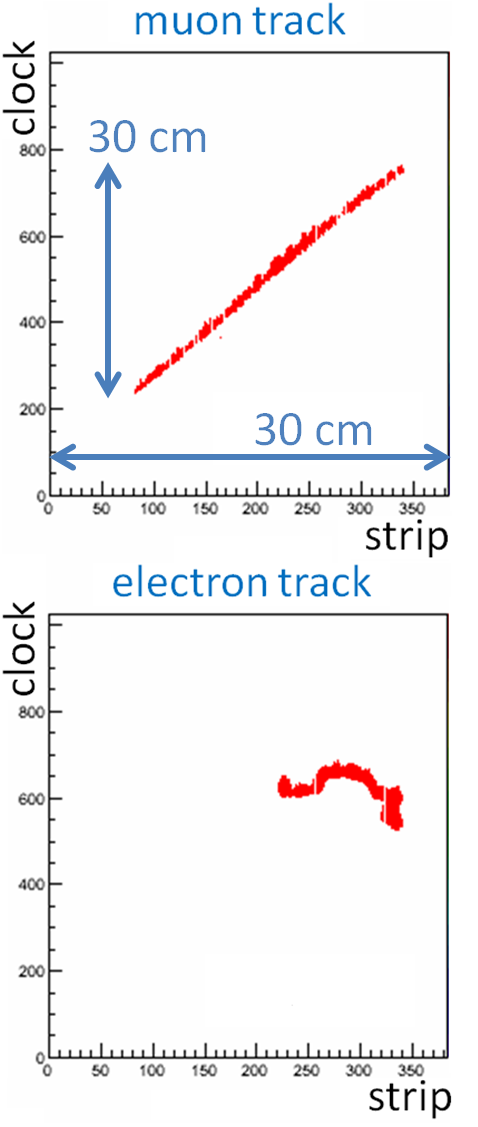} &
    \includegraphics[width=.35\textwidth]{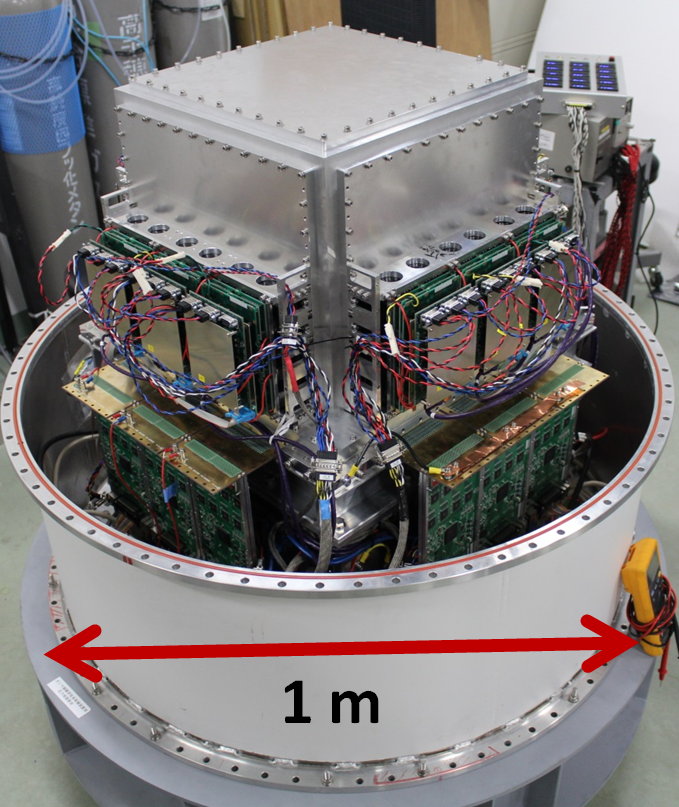}
  \end{tabular}
  \caption{Schematic view of an ETCC and the definition of $\alpha$ angle
           are illustrated in the left panel.
           Example images of a muon track and an electron track
           are drawn in the center panels.
           A photograph of SMILE-II's 30~cm-cube ETCC in
           the bottom half cup of the pressure vessel is shown in the right panel.}
  \label{fig_ETCC}
\end{figure}

An electron-tracking Compton camera (ETCC),
which is shown in the left panel of Fig.~\ref{fig_ETCC},
consists of a gaseous Time Projection Chamber (TPC)
and pixel scintillator arrays (PSAs).
An incident gamma ray is Compton scattered by 
an orbital electron of a gas molecule in the TPC,
and the TPC measures the track of the Compton recoil electron and its energy.
The TPC has a two-directional strip-readout type
position sensor known as the $\mu$-PIC~\cite{ochi-01}
which measures lateral (x and y-axis) information of particle tracks
amplified by a gas electron multiplier (GEM~\cite{sauli-97}),
and the remaining vertical depth (z-axis) is determined by
the clock count of the signals of down-drifting ionized electron cloud.
The track of a charged particle is obtained as
two two-dimensional projected images
which reflect hit strips and depth information.
Examples of a measured muon and electron track are shown
in the center panels of Fig.~\ref{fig_ETCC}.
The Compton scattering position and
the initial direction of the recoil electron
are obtained by analysis of the track.
In addition,
the Compton scattered gamma ray is absorbed by the PSAs,
which measure the absorption position and the energy.
The combination of TPC and PSA data
enables us to measure all of the parameters of the Compton scattering kinematics,
allowing both the energy and the arrival direction of
an incident gamma ray to be reconstructed using the ETCC.
 
\begin{figure}[tbp]
  \centering
  \includegraphics[width=.75\textwidth]{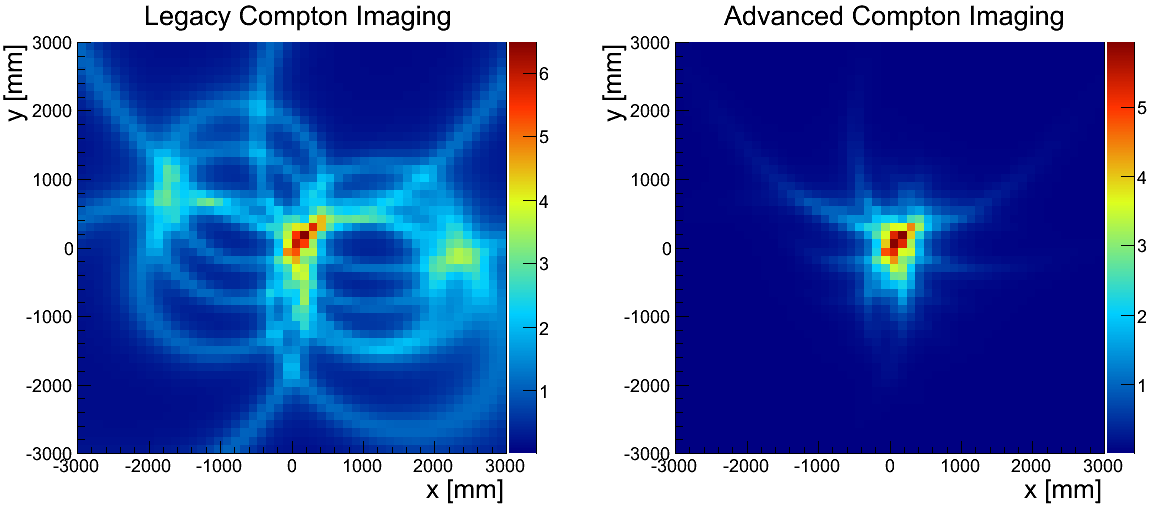}
  \caption{Comparison of imaging of a few gamma rays with legacy
           and with advanced Compton techniques.
           The left panel is the legacy imaging, where
           the reconstructed gamma rays are drawn as doughnut-shapes.
           The right panel is the advanced imaging, where
           the gamma rays are drawn as crescent-shapes.}
  \label{fig_leg_adv}
\end{figure}

Measurement of the electron track gives unique and
significant advantages to the Compton camera.
At first, the reconstructed gamma-ray direction is determined as
a point in the sky.
The shape of directional uncertainty of an event 
is constrained to a crescent shape.
This shape is made by a combination of 
the resolution of the Compton scattering angle
(angular resolution measure, ARM)
and directional angular resolution of the Compton scatter plane
(scatter plane deviation, SPD).
In the case of a COMPTEL-type legacy Compton camera
which has no electron track information,
the incident gamma ray is reconstructed as a ring on the sky,
with a doughnut-shaped directional uncertainty.
Examples of imaging with and without electron tracks
are shown in Fig.~\ref{fig_leg_adv}.
There are some artifact peaks in the legacy Compton image,
while such artifacts disappear with our advanced Compton imaging technique.
Furthermore, 
the combination of the pulse height and the range of track information
enables us to measure
the energy loss rate (dE/dx) of charged particles in the TPC, 
which provides efficient background rejection.
By using dE/dx,
fully-contained electrons stopping in the TPC are discriminated 
from other charged particles such as 
MIP-like high energy electrons, cosmic muons, protons, and so on.
There also remains one unused measured parameter which can be
used for background rejection or improvement of reconstruction accuracy.
For background rejection,
the $\alpha$ angle between the directions of the recoil electron and 
the scattered gamma ray is used to check the Compton scattering kinematics.
This check rejects non-Compton scattering events such as 
chance coincidence hits of the TPC and PSAs,
and events where Compton scattering occurred in the PSAs.
Background rejection techniques using track information,
dE/dx and $\alpha$ angle selection,
are a unique method for effective background suppression 
for observations of continuum gamma-ray sources.
For the improvement of reconstruction accuracy,
we can search the most likely set of measured parameters
in the ranges of their uncertainty,
but this topic is beyond the scope of this article.

\subsection{Status of the current SMILE-II ETCC}

We have developed a 30~cm-cube size ETCC as
the flight model detector of SMILE-II.
The right panel of Fig.~\ref{fig_ETCC}
is a photograph of the flight model ETCC
in the bottom half cup of the pressure vessel.
The ETCC consists of the 30~cm-cube TPC and 108 PSAs of GSO crystal.
The TPC and 36 PSAs placed under the bottom of the TPC
have been installed to the system.
The other 72 PSAs around the four lateral sides of the TPC
will be installed in the system soon.
To improve the data quality of the electron track, dead time ratio,
power consumption, and overall weight of the instrument,
we have fully replaced the DAQ system from SMILE-I~\cite{takada-11}
and the prototype 30~cm-cube size ETCC~\cite{ueno-12}.
Current detector parameters are summarized in the Table~\ref{tab_TPC}.
\begin{table}[tbp]
  \centering
  \caption{Current parameters of the SMILE-II ETCC}
  \label{tab_TPC}
  \begin{tabular}{|ccc|}
    \hline 
    \multirow{5}{*}{TPC}
    & TPC size                    & $30\times30\times30$~cm$^3$ \\
    & Gas                         & Ar:iso-C$_4$H$_{10}$:CF$_4$ = (95:2:3), 1~atm.\\
    & Drift velocity              & $\sim 6$~cm/$\mu$s \\
    & $\mu$-PIC readout fineness  & 800~$\mu$m pitch, 100 MHz frequency \\
    & Pulse height sampling       & 50 MHz FADC (Sum of 32 strips) \\
    & Energy resolution (FWHM)    & 22 \% (@ 22~keV) \\
    \hline
    \multirow{4}{*}{PSAs} 
    & Scintillator             & GSO:Ce (6.71 g/cm$^3$) \\
    & Pixel size               & $6\times6\times13$~mm$^3$ \\
    & Dynamic range            & 80 keV--1.3 MeV \\
    & Energy resolution (FWHM) & 10 \% (@ 662~keV) \\
    \hline 
  \end{tabular}
\end{table}
\begin{figure}[tbp]
  \centering
  \includegraphics[width=.95\textwidth]{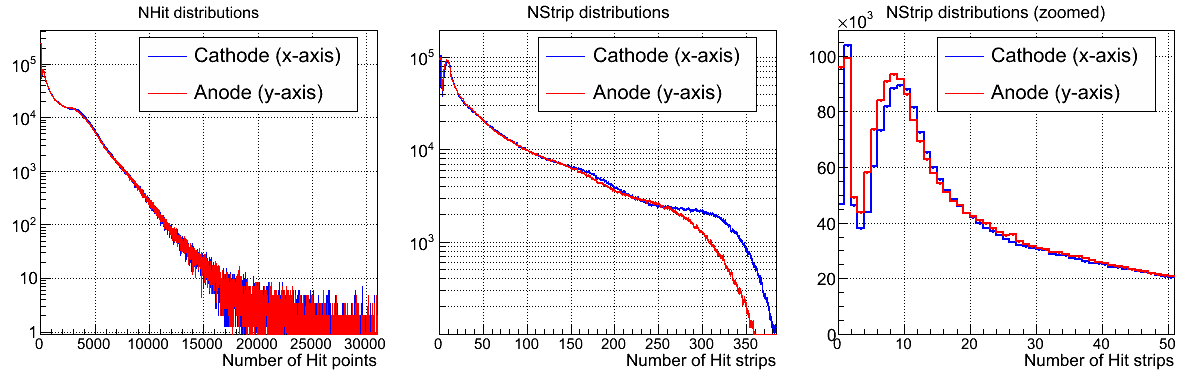}
  \caption{Number of hit points spectra (left),
           and number of hit strips spectra (center and right).
           The red curves are anode hits,
           and the blue curves are cathode hits.
           Number of hit strips have two peaks,
           the first component is chance coincidence noise,
           and the other is the signal component with peak around 10 strips.}
  \label{fig_nhit}
\end{figure}
As a result of the DAQ upgrade,
the quality of the tracks is dramatically improved,
and the number of hit points and strips in the obtained track images
are increased by more than one order of magnitude compared to the old systems.
These hit number spectra are shown in Fig.~\ref{fig_nhit}.
Details of the current DAQ concept and hardware appear in
S.~Komura~et~al.~\cite{komura-13} and 
T.~Mizumoto~et~al.~\cite{mizumoto-13}.

\begin{figure}[tbp]
  \centering
  \begin{tabular}{cc}
    \includegraphics[width=.40\textwidth]{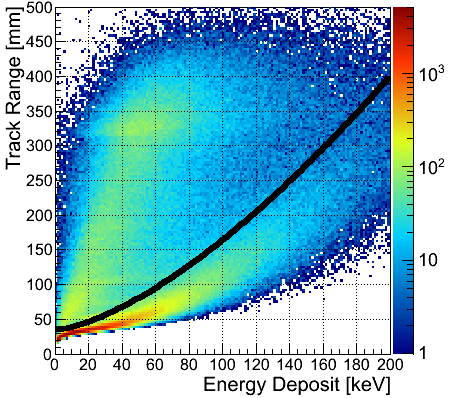} &
    \includegraphics[width=.45\textwidth]{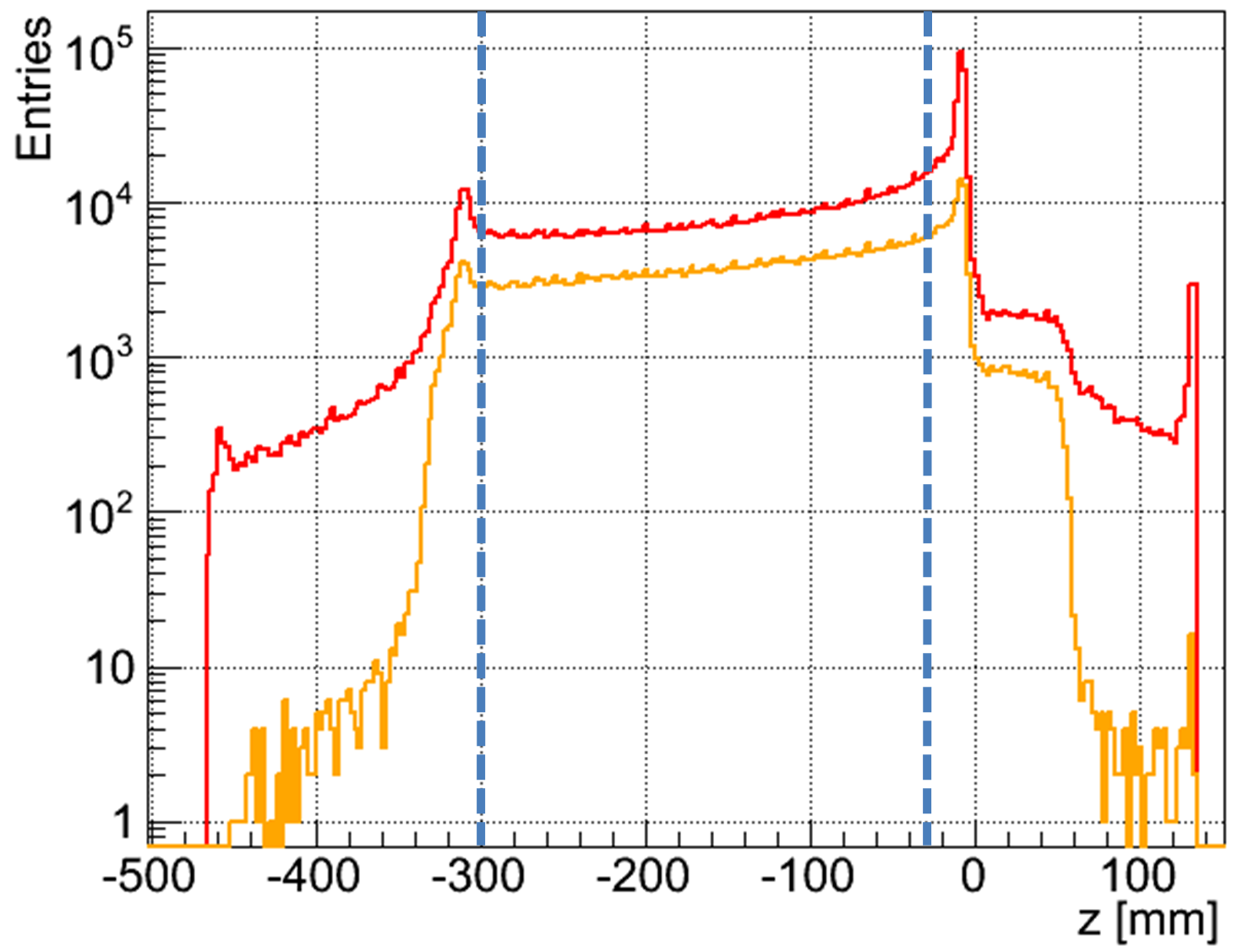}\\
  \end{tabular}
  \caption{The left panel shows Energy loss rate (dE/dx) map
           of charged particles in the TPC.
           It is the relation of the track range
           and energy deposit of charged particles.
           The solid line indicates dE/dx cut criteria for
           the discrimination of fully-contained electron
           from other charged particles in this analysis.
           The right panel is the distribution of the
           The distributions of Compton scattering position
           along the z-axis (TPC clock) direction,
           where the red and orange curve indicates before and after
           the dE/dx cut, respectively.
           Two vertical dashed blue lines indicate
           the fiducial volume region
           used in this analysis.}
  \label{fig_dEdx}
\end{figure}

\subsection{Simple analysis of electron track}
The Compton scattering position and direction of the recoil electron
are determined by track analysis of the TPC data.
The track data are obtained as two two-dimensional images
whose dimensional spaces are
TPC clock (z-axis) versus $\mu$-PIC strips (x or y-axis).
At first, we reconstruct the three dimensional track from these two track images
by simple off-line coincidence of the images in three dimensional space.
Next, we make an image by projection of
the obtained three dimensional track to the x-y plane.
As the most simple analysis,
we consider that the Compton scattering occurred in
the hit pixel of the x-y image which is closest to
the photo-absorbed position~\cite{komura-13}.
We also adopt the mean of hit clocks in this closest pixel as
the z-axis position of the Compton scattering.
The direction of the recoil electron is determined as
a composite vector of the two gradients of the obtained track images,
which are the x-z or y-z gradients of each of the two dimensional tracks.
It is a rough track analysis
for a quick test of the first performance of the ETCC.

A measured dE/dx map is shown in Fig.~\ref{fig_dEdx}.
The map is beautifully divided into two components around the solid line.
One is the fully-contained electrons stopping in the TPC
which events are distributed in the high dE/dx area of the map.
Another component is MIP-like charged particles,
such as cosmic muons and high energy electrons escaping from the TPC.
For the background rejection,
we use the dE/dx cut criteria as following:
\begin{eqnarray}
   \left( \frac{\rm Track~Range}{\rm [mm]} \right)
   > 4.1\times10^{3} \left( \frac{K_{\rm e}}{\rm [MeV]} \right)^{1.5} + 35,
\end{eqnarray}
where $K_{\rm e}$ is the energy deposit of a charged particle in the TPC,
and the track range is determined by
geometrical combination of the two two-dimensional tracks.
This dE/dx criteria is drawn as the solid line
in the left panel of Fig.~\ref{fig_dEdx}.

The analyzed starting positions of the measured tracks
(i.e., the Compton scattering position) along the z-axis,
both before and after the dE/dx selection, are
drawn in the right panel of Fig.~\ref{fig_dEdx}.
The drift plane, GEM, and $\mu$-PIC of the TPC are located at
$z = -316.9$, $-7.9$, and $-2.9$~mm, respectively.
Events scattered on the inside of the TPC are reduced by
about a factor three by the dE/dx cut.
Otherwise, events scattered outside of the TPC are greatly reduced by
about two orders of magnitude.
The latter is considered as chance coincidence noise,
therefore the dE/dx selection has efficient background rejection capability.
In addition,
we defined the fiducial volume region as $-300~{\rm mm} < z < -30~{\rm mm}$
for valid Compton scattering position in the following analysis.
We also use x-axis and y-axis fiducial cuts
to avoid distortion of electrical field in the TPC.
Thus, the current aperture ratio of the x-y fiducial area
to the full $\mu$-PIC size is about 0.57.
Although this loss has a large effect for current performance of detector,
the principal cause has already been identified as being 
due to the positions of the support posts of the drift-cage,
and will be fixed soon by changing the post positions.
Moreover, we cut events with a reconstructed direction of
the zenith angle larger than 90 degrees.
Details of the reconstruction method of direction
is written in A.~Takada~et~al.~\cite{takada-11}.


\section{Results of performance test}

Here, we show results of the first performance test
of the 30~cm-cube ETCC by using gamma-ray sources of
$^{139}$Ce (166~keV), $^{133}$Ba (356~keV), $^{22}$Na (511~keV),
and $^{137}$Cs (662~keV).
%
%

\begin{figure}[tbp]
  \centering
  \includegraphics[width=.99\textwidth]{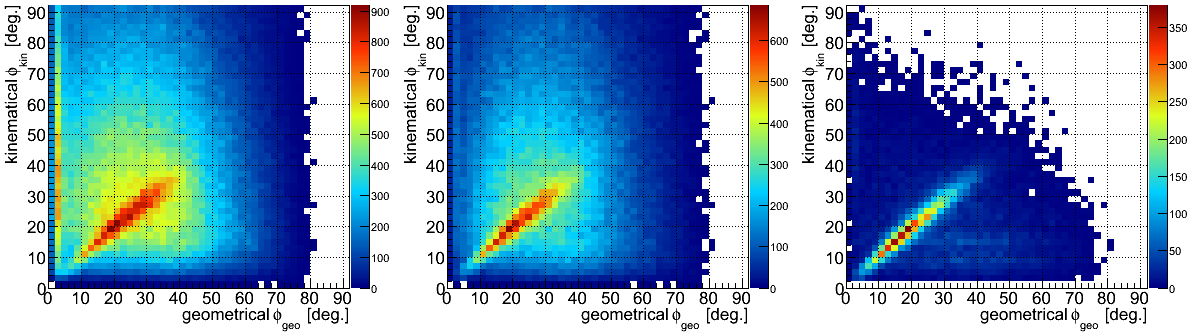}
  \caption{Relation of Compton scattering angle $\phi$,
           determined both kinematically ($\phi_{\rm kin}$) and
           geometrically ($\phi_{\rm geo}$),
           under 662~keV gamma-ray irradiation.
           The left and center panels are before and after
           the dE/dx cut, while the right panel is includes also
           the energy selection of 662~keV $\pm10\%$.}
  \label{fig_phi}
\end{figure}
Fig.~\ref{fig_phi} shows the relation maps of the Compton scattering
angle $\phi$, determined both kinematically and geometrically,
under 662~keV gamma-ray irradiation.
The kinematical scattering angle $\phi_{\rm kin}$
is derived from the sum and ratio of energies of the recoil electron and
the scattered gamma ray.
The geometrical scattering angle $\phi_{\rm geo}$ is determined from
the recoil electron direction and Compton scattering position
with the assumption of the gamma-ray source direction.
The appearance of the correlation in each of these maps
is evidence of the detection of source-derived gamma rays.
Moreover, the background component is
clearly suppressed by the dE/dx selection.
By using the energy selection window of 662~keV~$\pm10\%$,
environmental room gamma-ray backgrounds and
the component scattering outside of the TPC are suppressed.

\begin{figure}[tbp]
  \centering
  \includegraphics[width=.78\textwidth]{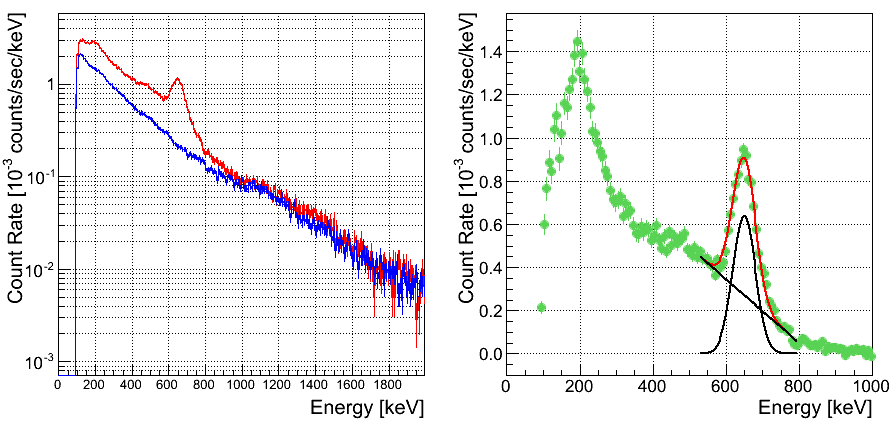}
  \caption{Observed energy spectra of the current SMILE-II 30~cm-cube ETCC.
           Livetime-normalized spectra are shown in the left panel, where
           the red line is under 662~keV gamma-ray irradiation,
           and the blue line is the background spectrum.
           The green points in the right panel
           indicate the excess spectrum.
           The red curve is a fit function consisting of
           a Gaussian and a linear equation.}
  \label{fig_spect}
\end{figure}

\begin{figure}[tbp]
  \centering
  \includegraphics[width=.99\textwidth]{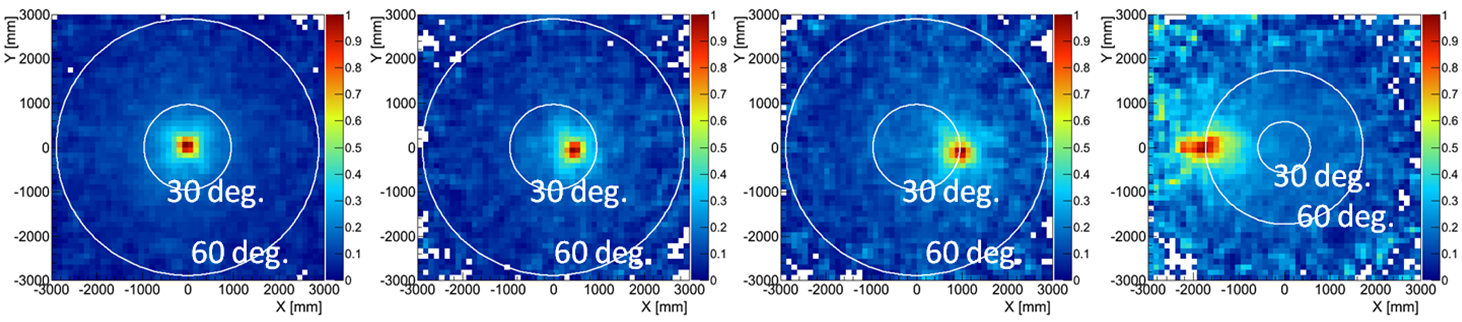}
  \caption{Simple backprojection images of a $^{137}$Cs gamma-ray source
           at zenith angles of 0, 15, 30, and 60 degrees.
           The white rings indicate 30 and 60 degrees from
           the center of the field of view.
           The $^{137}$Cs source is found at the expected angle
           in the all of these images.
           Therefore, the ETCC covers a field of view wider than 3 sr.}
  \label{fig_bkproj_maps}
\end{figure}

The Observed energy spectra after event selections are shown in Fig.~\ref{fig_spect}.
The line peak of 662~keV gamma rays from the $^{137}$Cs is found in
the background-subtracted spectrum.
The energy resolution is measured to about $11\%$ at 662~keV,
which is defined as FWHM of the fitted Gaussian function.
Simple backprojection images of the $^{137}$Cs gamma-ray source
at zenith angles of 0, 15, 30, and 60 degrees
are drawn in Fig.~\ref{fig_bkproj_maps}.
The source is found at the expected position in the all of these images.
It shows that the ETCC has a field of view wider than $3~{\rm sr}$.

Angular resolution parameters of the Compton camera,
ARM and SPD distributions,
are also evaluated by using gamma-ray checking sources.
The ARM distribution measured with the 662~keV gamma-ray source
at the center of field of view is well explained by
a Lorentzian function with FWHM of 5.3 degrees.
The SPD distribution under same conditions as the above with
the additional event selection
of $|{\rm ARM}| < 5.3/2$ degrees is explained by
a combination of a Gaussian function with FWHM of 200 degrees and a constant.
The primary peak component of the point spread function (PSF)
is dominated by the ARM resolution,
while the SPD resolution dominates the outer tail component of the PSF.
This good ARM resolution fulfills the requirement
of less than 10 degrees for SMILE-II.

We also carried out a test of gamma-ray detection efficiency
of the current SMILE-II ETCC,
and we have achieved efficiencies on
the order of $10^{-4}$ for the sub-MeV energy range.
This good performance is about one order higher than the previous 
SMILE-I type ETCC.
Furthermore, our obtained efficiencies for a few hundred keV band
are consistent within $10\%$ with
the simulation which takes into account only physical processes~\cite{sawano-13}.
It means almost all of the fully-contained Compton recoil electrons
are detected by the TPC.
Thus, our simulation of performance enhancements for upgrades of the detector,
such as scintillator materials and its radiation length,
layout design of PSAs,
fiducial volume of the TPC,
gas blending ratio and its pressure in the TPC,
can be considered more reliable.


\section{Summary}
We have developed the SMILE-II flight model ETCC with
30~cm-cube size TPC,
and we have carried out the first performance test
at the room condition by using several gamma-ray sources
in the sub-MeV energy range.
Now, we use a simple analysis of the track
for a quick test of the performance.
The gamma-ray imaging capability is demonstrated
by clear images of gamma-ray sources, even at the edge region of
the field of view at 60 degrees from the center.
Therefore, we confirmed the width of the field of view as about 3~sr.
The angular resolution, FWHM of ARM distribution,
is determined as 5.3 degrees for 662~keV gamma rays.
Gamma-ray detection efficiencies are achieved to
the order of $10^{-4}$ in the sub-MeV energy range.
This good performance is about one order of magnitude better than
the previous SMILE-I type ETCC.
Good consistency of efficiencies between this performance test
and physical process simulation is very important as 
we obtain a dependable prediction method
of performance enhancements of detector upgrades.


With the combination of efficiencies and angular resolutions obtained here,
the Crab Nebula or Cygnus X-1 can be detected by our ETCC with a significance
above the $5\sigma$ statistical level in several hours observation in a 
one day balloon flight experiment at middle latitude.
We are going to confirm the sub-MeV gamma-ray imaging capability of the ETCC
for celestial objects by the SMILE-II experiment.
The SMILE-II experiment and following SMILE-III project
will be an important key to the progress of sub-MeV/MeV gamma-ray astronomy.

%
%

\acknowledgments

This work was supported by a Grant-in-Aid in Scientific Research from
the Japan Ministry of Education, Culture, Science, Sports and Technology.
This work was also supported by research fellowships for
young scientists from the Japan Society for the Promotion of Science (JSPS).
Some of the electronics development was supported by
the Open-It consortium.

\end{document}